# A Wireless System Using Random Residue Sequences

Vamsi Sashank Kotagiri
Oklahoma State University, Stillwater

**Abstract**
This paper describes the architecture of wireless communication system using random residue sequences. The basic scheme is that of spread spectrum but instead of using PN sequences for coding, we use random residue sequences. Such a system can provide cryptographic security whose strength would depend on the number of code sequences being used.

**Introduction**
Spread spectrum systems which have a variety of applications for secure communication use PN sequence-based codes [1]. These sequences provide near-ideal autocorrelation function but the period is constrained to be $2^n-1$, for different values of n [2]. The autocorrelation function, which captures the correlation of data with itself, is one of the measures of randomness if it is close to zero or zero for non-zero shifts. For a data sequence *a(n)* of *N* points the autocorrelation function *C(k)* is represented by

$$C_a(k) = \frac{1}{N}\sum_{j=1}^{N} a(j)a(j+k) \qquad (1)$$

For a noise sequence, the autocorrelation function $C_a(k) = E(a(i)a(i+k))$ is two-valued, with value of 1 for k=0 and a value approaching zero for k≠0 for a zero-mean random variable. Assuming periodicity, such a sequence will have C(k) as 1 for k=0 and approximately $\mu^2$ for non-zero *k* where *µ* is the mean of the variable [3]. There are other considerations related to frequency of subsequences [2] as well as testing with respect to a combination of criteria [4] that are employed to determine if a sequence is random.

Another interesting random sequence family is that of "decimal" sequences [5]-[9], which is of particular interest since any sequence can be described also as a "decimal sequence". If judiciously used, it can provide a large family of near-orthogonal sequences. However, due to the residual non-zero autocorrelation value for non-zero shifts, it will have some crosstalk if the same sequence is used by all the users. More recently, the number theoretic Hilbert transform (NHT) based sequences [10]-[12] have been proposed due to their *ideal* autocorrelation properties. We call them RR (random residue) sequences.

The NHT [10] as a generalization of the discrete Hilbert transform (DHT) [13], is a circulant matrix with alternating entries of each row being zero and non-zero numbers and transpose modulo a suitable number is its inverse. The DHT has found many applications in signal processing and also in scrambling and in a variety of other applications in speech and image analysis [14]-[18] and it is closely related to scrambling transformations [19]-[21].

Let us consider the data block to be F and the NHT transform to be N. The matrix N is the general form of NHT and m is appropriate value of modulus; its inverse is $N^T$ mod q. For a block of data F, the NHT transform $G = NF$ mod $q$. The inverse of the transform is $F = N^T G$ mod $q$. In other words,

$$NN^T = I \qquad (1)$$

where I is the identity matrix.

**Spread Spectrum based communication system**
The main idea behind spread spectrum was to use more bandwidth than the original message by maintaining same signal power. As the spread spectrum signal does not have a clear distinguishable peak in the time domain this makes the signal difficult to distinguish from its noise. Spread spectrum is used to provide secure communication by spreading the given signal over a large frequency band, because of this reason spread spectrum signals can transmit with low spectral power density.

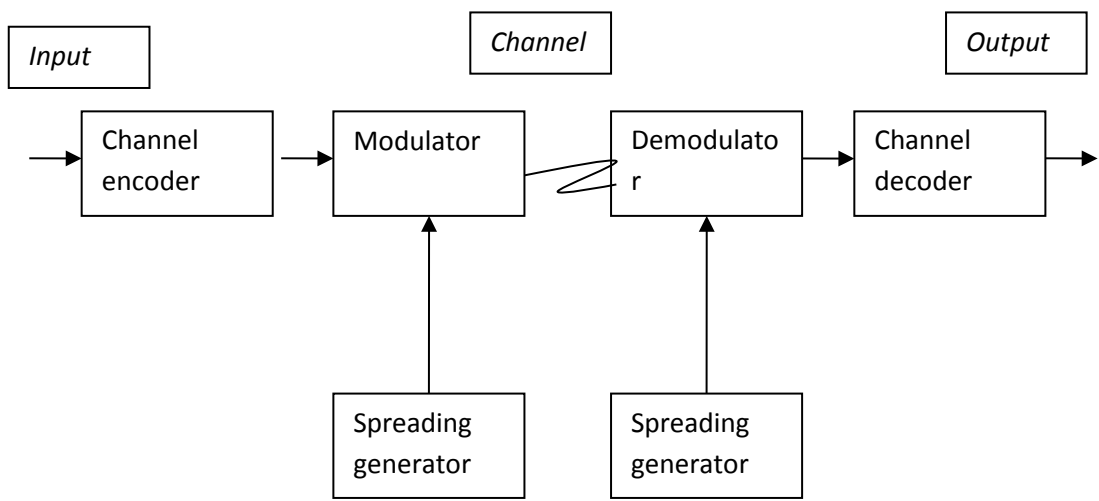

Figure 1: Spread spectrum communication system

A large number of users want to share a common channel to transmit information to a receiver in a multiple access system scenario. As the system has a fixed amount of resources, spectrum and channels, the system has to manage resources appropriately as multiple units are trying to access the system at the same time. To solve this problem three methods are used: frequency-division multiple access (FDMA), time-division multiple access (TDMA), code-division multiple access (CDMA). Both TDMA and FDMA have drawbacks and they are inefficient for multiple access system limits, we need to find an alternative i.e. by allowing more than one user to share a channel by use of direct sequence spread spectrum signals (DS-SS). Wireless standard IEEE 802.11 uses either FHSS or DSSS in its radio interface [22]. For a background of CDMA schemes, see [23].

In Figure 1, each user is assigned a unique code sequence that allows the user to spread the information signal across the assigned frequency band. Signals from the various users are separated at the receiver by cross-correlation of the received signal with each of user code sequences. The cross-correlation and the cross-talk inherent in demodulated signals received are minimized as a result of designing the code sequences. The above multiple access method is CDMA, in order to classify a system as SS modulation technique, the transmission bandwidth must be much larger than the information bandwidth and the resulting RF bandwidth must be determined by a function other than the information being sent.

The intentional and unintentional interference and jamming signals are rejected because they do not contain the spread spectrum key. Only the desired signal which has the key will be seen at the receiver while demodulating the received signal.

**System architecture using RR sequences**

A wireless network system consists of several components that support communications by converting information signals into suitable form for transmission through the wireless medium. Wireless networks include computer devices, base stations and a wireless infrastructure. The user is the party that initiates and terminates the information signal in wireless networks and can directly utilize the wireless network.

In our proposed architecture we have N users, where each user wants to transmit their message or information signal securely without being intercepted by other user's signals and only the authentic user will be able to receive his information signal. The first user will be having his message $M_1$ to transmit, he will mix message $M_1$ with a random residue sequence $V_1(t)$ before transmission. Same procedure will be followed by the remaining users who wish to transmit their message signal, but each user will be getting a random left or right shift of the original random residue sequence, which each user will be mixing to his message before transmission.

At the receiver end, if a user wants to get back his original message or information signal the he has to mix his own random residue sequence to the message signal received. We know [11],[12] the autocorrelation of the random residue sequence for all non-diagonal shifts is zero, so only the authentic user who is having a correct RR sequence will be able to retrieve original message signal. Similarly the remaining users can retrieve their original message by using their valid RR sequences. Thus only the authentic users can retrieve their message signal and all intruders cannot get access to the original message which also provides security to the information signal.

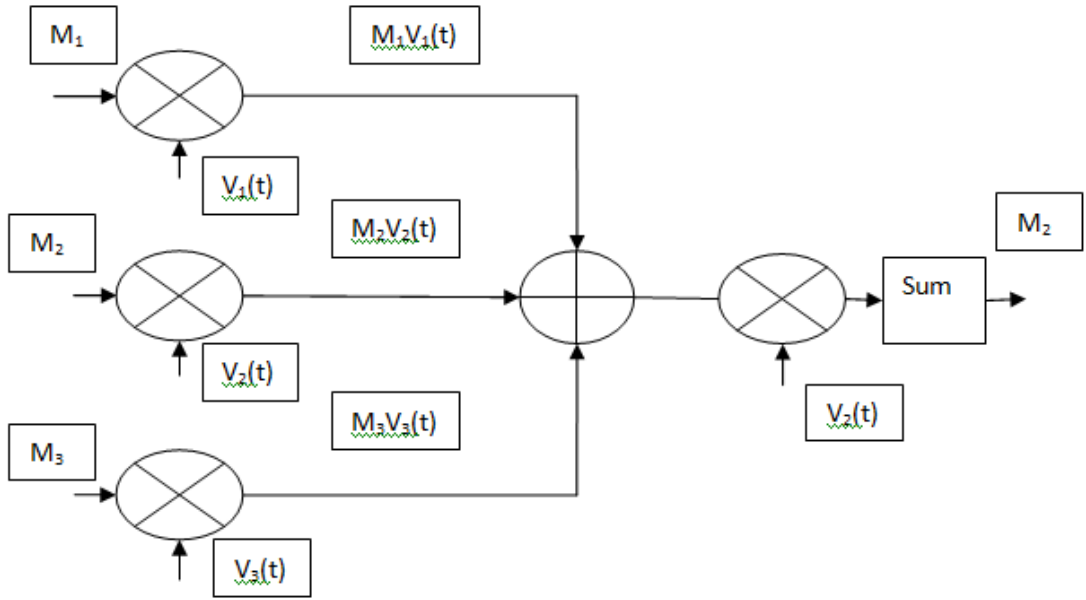

Figure 2: Wireless system communication architecture using random residue sequences

**Implementation of wireless system architecture using a 16 bit RR sequence**
The wireless system will be illustrated by an example. Let the 16 bit sequence be a=11, b=2, c=4, d=8, e=16, f=32, g=17, h=34, i=21, j=42, k=37, l=27, m=7, n=14, o=28, p=9. In a multiple access system, more than one user wants to communicate through the same common channel.

As the users share the same common channel to transmit their message, each user is assigned unique RR sequence i.e. the user $M_1$ will be assigned the original sequence 11, 2, 4, 8, 16, 32, 17, 34, 21, 42, 37, 27, 7, 14, 28, 9 the user $M_2$ sequence might be left shift or right shift of the original sequence 14, 28, 9,11, 2, 4, 8, 16, 32, 17, 34, 21, 42, 37, 27, 7 and the user $M_3$ sequence is 4, 8, 16, 32, 17, 34, 21, 42, 37, 27, 7, 14, 28, 9, 11, 2 which is 2 bits left shift of the original sequence.

The combined signal along with user's unique rr sequence will be transmitted and the signal will be separated at the receiver by autocorrelation with each user unique rr sequence.

User $M_1$ rr sequence: 11, 2, 4, 8, 16, 32, 17, 34, 21, 42, 37, 27, 7, 14, 28, 9

User M$_2$ rr sequence: 14, 28, 9, 11, 2, 4, 8, 16, 32, 17, 34, 21, 42, 37, 27

User M$_3$ rr sequence: 4, 8, 16, 32, 17, 34, 21, 42, 37, 27, 7, 14, 28, 9, 11, 2

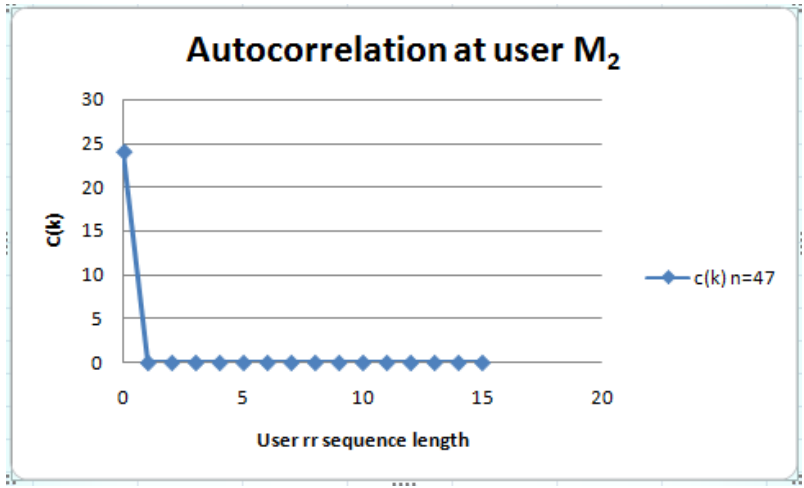

Figure 3: Zero autocorrelation produced when user M$_2$ produces his sequence at receiver

As each user will be having his own RR sequence, so the crosstalk inherent in the demodulated signals received will be minimized.

**Conclusions**

This paper describes the architecture of wireless communication system using random residue sequences. The basic scheme is that of spread spectrum but instead of using PN sequences for coding, we use random residue sequences. Such a system can provide cryptographic security whose strength would depend on the number of code sequences being used.